# Digital Equalization of Ultrafast Data Using Real-time Burst Sampling

**Ali Motafakker-Fard, Shalabh Gupta, and Bahram Jalali**
*Department of Electrical Engineering, University of California, Los Angeles*
*Author email address: motafakk@ee.ucla.edu*

**Abstract:** We demonstrate *digital* equalization of 40Gbps data enabled by use of *Real-time Burst Sampling (RBS)* with photonic time-stretch A/D converter. Evaluation of a feed-forward equalizer and a decision feedback equalizer is also shown.

**1. Introduction**

To satisfy increasing internet traffic demands, higher data rate optical communications links are required. Data rates at 40 Gbit/s and 100 Gbit/s using advanced modulation formats such as DQPSK and polarization multiplexing have become possible and are actively pursued [1-3]. To demodulate these signals, receivers will increasingly need to rely on A/D conversion and digital signal processing at the receiver, in much the same way as radio receivers operate today. However, bandwidth limitations of electronic A/D converters represent the major bottleneck in such links.

The photonic time-stretch A/D converter (TS-ADC) [4, 5] can provide continuous digitization of ultra-high bandwidth electronic signal by exploiting multiple parallel wavelength channels and extending the bandwidth of electronic converters [6]. Furthermore, the TS-ADC can provide better resolution than purely electronic ADCs [7], which is required for recovering data from optical links that use advanced modulation formats and polarization multiplexing. The Time-Stretch Enhanced Recording (TiSER) oscilloscope, which is the single channel version of the TS-ADC, uses photonic time stretch pre-processing to perform *Real-time Burst Sampling (RBS)* of ultra-wideband signals [8]. It is particularly well suited as a diagnostic and development tool for high data rate communication systems because of its ability to capture bursts of samples, spanning several data symbols, in real-time. This instrument can be very useful in studying different equalization techniques, and therefore can reduce costly design iterations that are required for development of optical transceivers [9, 10].

In this paper we present proof-of-concept results showing digital equalization of 40Gbps non-return-to-zero on-off keying (NRZ-OOK) data captured using the TiSER oscilloscope. The scope captures 40 Gbps data by combining a photonic time stretch pre-processor and an electronic digitizer with a mere 1 GHz input bandwidth. Digital equalization is enabled by the ability of the TiSER to capture bursts of ultrafast data in real-time, an important function that is absent in a sampling oscilloscope. We also discuss the comparison between feedforward equalization and decision feedback equalization at such data rates.

**2. Real-time burst sampling with photonic time-stretch digitizer**

Although equivalent-time sampling oscilloscopes can capture electrical signals with up to 100 GHz bandwidth, fast non-repetitive dynamics and rare events cannot be captured with these instruments. More importantly, information about intersymbol interference (ISI) in random or pseudo-random data transmission is not provided by a sampling scope because it lacks real-time capability. Yet such information is required for correction of channel impairments (such as dispersion and optical nonlinearities) using digital signal processing [11, 12] under true, i.e. random or pseudo-random data transmission. On the other hand, real-time electronic digitizers can capture electrical signals continuously but their bandwidth is limited up to 20 GHz for available systems (with about 5-6 effective number of bits). Hence, they don't meet the requirements of 40Gbps NRZ-OOK and the upcoming 100Gbps optical links..

The real-time optical sampling system demonstrated in [13] overcomes these limitations and can capture signals with 50 GHz bandwidth in real-time. However, this system requires an optical input as it relies on all-optical sampling. Also, the implementation is complicated as it requires not one, but a parallel channels of very high speed digitizer, and complex signal processing to remove interchannel crosstalk.

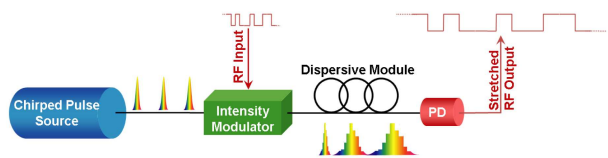

**Fig. 1:** Time stretch preprocessor. The system is a modified dispersive optical link that uses a chirped supercontinuum source and high dispersion, low loss, fiber.

While TiSER operates in the equivalent-time mode, it is superior to a sampling oscilloscope because, at each sampling period, it captures not just one, but rather bursts of samples in real-time. Hence, TiSER offers a new mode of sampling called *Real-time Burst Sampling (RBS)*. This sampling mode enables capture of fast non-repetitive dynamics and rare events that occur at or near the modulation rate. These deviations can not be captured with equivalent-time oscilloscopes because they lack real-time capability, or with conventional real-time digitizers because of insufficient bandwidth [14].

In time-stretch analog-to-digital converter (TS-ADC), the signal is modulated over linearly chirped optical pulses (Fig. 1). These chirped pulses can be obtained by dispersing ultra-short pulses from a mode-locked laser (MLL) or a super-continuum source. Propagation through a dispersive medium, such as dispersion compensating module, stretches RF modulation. As a result, the very large bandwidth signals can be captured in real-time by a much slower electronic digitizer. If problematic, the dispersion induced bandwidth limitations can be entirely eliminated using phase diversity [15], single side band modulation [16] or digital backpropagation [17]. The bandwidth is then limited only by the electro-optic modulator, which is presently up to 100 GHz for commercial devices, such as those manufactured by GigOptix [18].

## 3. Digital Equalization

The TiSER oscilloscope used here consists of a mode-locked laser generating ultra-short pulses at 36 MHz repetition rate, followed by a -20 ps/nm dispersion compensating fiber giving chirped pulses at its output. A Mach-Zehnder intensity modulator is used for modulating the RF signal over these chirped pulses. Another fiber with dispersion value of -1310 ps/nm stretches the modulated optical pulses in time, resulting in a stretch factor of 67. Finally, a photo-detector converts these pulses back to electrical domain generating a signal that contains stretched replica of the original RF signal segments. The RF signal's spectrum is now modified in such a way that both the carrier frequency and bandwidth are reduced. A commercial digitizer with 1 GHz bandwidth and 50GS/s sampling rate is used for capturing this signal. In order to achieve an integer number of samples per data symbol, the captured signal is down-sampled to 2.39 GS/s in digital domain to represent 4 samples per data symbol. The laser and data clocks are recovered digitally and an eye diagram is generated.

The eye diagram is constructed by removing integral number of data periods from the stretched time scale to overlay these time segments. The functional block diagram for the equalization experiment is shown in Fig. 2.

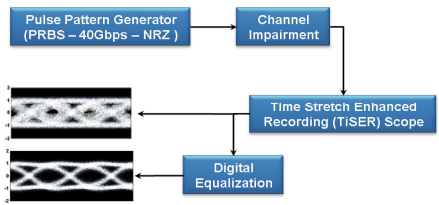
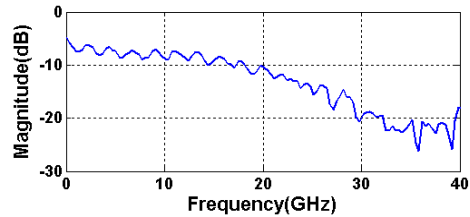

**Fig. 2:** Block diagram of the experimental setup.  **Fig. 3:** Channel frequency response

Passive electrical components with impedance mismatches were used to emulate impairments in the form of ripples in the channel response. Furthermore, the 20-GHz modulator was used along with 40 Gbps data to emulate the low pass characteristics of the channel. The frequency response of the impaired channel is shown in Fig. 3.

Equalization techniques use a multi-tap feedforward (FFE) or a decision feedback (DFE) filter to compensate for channel distortion [19]. In the present configuration the TiSER captures, in real time, segments spanning 9 symbols. A 4-tap FFE and a 4-tap DFE with tap spacing of about 0.25 symbol period are used with adaptive LMS algorithm to equalize the 40Gbps data eye distorted by the channel response. Here the 4-tap sampling window runs/scans across the real-time symbol stream captured by TiSER.

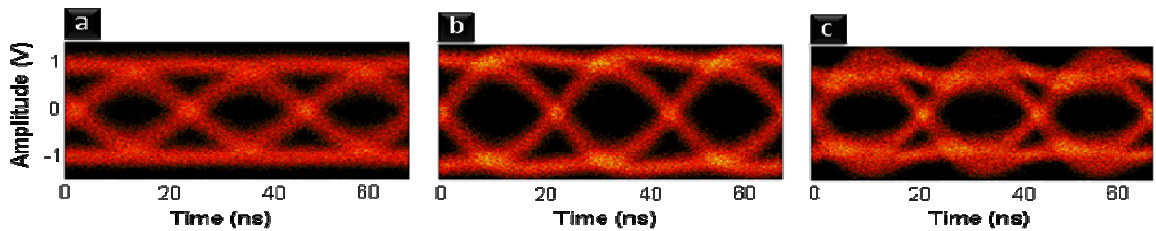

**Fig. 4:** Eye diagram for 40 Gbps NRZ-OOK data (a) before equalization (b) after feedforward equalization (c) after decision feedback equalization.

In order to obtain the equalizer weights in each case, the adaptive equalization algorithm is performed on the first few segments of a known sequence used as the training signal. Once the equalizer weights are determined, the equalizer is applied to other segments. If the channel characteristics may change over time, depending on the time scale of the changes, the adaptive algorithm may be run as needed to refresh the weights. Such is the requirement with all equalization techniques and is independent of the method of capturing the data. The results shown in Figure 4 show clear improvement in the equalized 40Gbps data compared to the raw received data. Figure 5 shows the normalized error of the equalized and distorted signal for each type of equalizations, i.e. FFE and DFE. This figure implies that

DFE converges with less error compared to FFE, resulting in better bit-error ratio (BER).

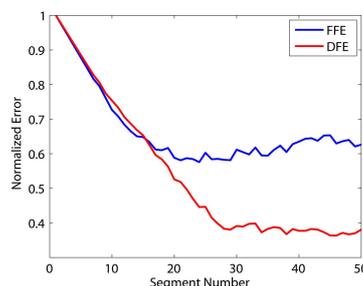

**Fig. 5:** Normalized error for decision feedback equalizer (DFE) and feedforward equalizer (FFE).

**4. Conclusion**

We have demonstrated digital equalization of 40Gbps NRZ-OOK data using the TiSER oscilloscope. We also evaluate two types of equalization techniques for compensation of the non-ideal channel response. With its ability to capture streams of high speed data in real time, TiSER is a valuable tool for rapid link characterization and development of equalization techniques in next generation fiber optic links that have bandwidths beyond the reach of any conventional real-time oscilloscope available today.

**5. Acknowledgment**

The authors would like to thank Centellax for lending their 40Gbps PRBS generator and LN modulator driver. This work was supported by funding from National Science Foundation through CIAN ERC under grant #EEC-0812072, and by DARPA.

**6. References**

[1] C. Yu, T. Luo, B. Zhang, Z. Pan, M. Adler, Y. Wang, J. E. McGeehan, and A. E. Willner, "Wavelength-Shift-Free 3R Regenerator for 40-Gb/s RZ System by Optical Parametric Amplification in Fiber," IEEE Photonics Technol. Lett., **18**, 2569-2571, 2006.

[2] X. Wu, L. Christen, S. Nuccio, O. F. Yilmaz, L. Paraschis, Y. K. Lize, A. E. Willner, "Experimental Synchronization Monitoring of I/Q Data and Pulse-Carving Temporal Misalignment for a Serial-Type 80-Gbit/s RZ-DQPSK Transmitter," in Optical Fiber Communication (OFC 2008), paper OTuG2.

[3] G. Raybon, P. J. Winzer, A. H. Gnauck, A. A. Adamiecki, D. A. Fishman, N. M. Denkin, Y. H. Kao, S. Scrudato, T. L. Downs, C. R. Doerr, T. Kawanishi, K. Higuma, Y. Painchaud, C. Paquet, "107-Gb/s Transmission over 700 km and One Intermediate ROADM Using LambdaXtreme® Transport System," in Optical Fiber Communication (OFC 2008), paper OMQ4.

[4] A. S. Bhushan, F. Coppinger, and B. Jalali, "Time-stretched analogue-to-digital conversion," Electron. Lett. **34**, 839-841 (1998).

[5] B. Jalali and F. Coppinger, "Data Conversion Using Time Manipulation," US Patent number 6,288,659, (2001).

[6] J. Chou, J. Conway, G. Sefler, G. Valley, B. Jalali, "150 GS/s real-time oscilloscope using a photonic front end," in Microwave Photonics 2008, pp.35-38 (2008).

[7] S. Gupta and B. Jalali, "Time-warp correction and calibration in photonic time-stretch analog-to-digital converter," Opt. Lett. **33**, 2674-2676 (2008).

[8] S. Gupta and B. Jalali, "Time stretch enhanced recording scope," Appl. Phys. Lett. **94**, 041105 (2009).

[9] A. Garg, A.C. Carusone, and S.P. Voinigwscu, "A 1-Tap 40-Gb/s Look-Ahead Decision Feedback Equalizer in 0.18-μm SiGe BiCMOS Technology," J. Solid State Circuits, **41**, 2224- 2232 (2006).

[10] B. Franz, F. Buchali, D. Rosener, and H. Bulow, "Adaptation techniques for electronic equalizers for the mitigation of time-variant distortions in 43Gbit/s optical transmission systems," in Optical Fiber Communication Conference (2007), paper OMG1.

[11] Ezra Ip and Joseph M. Kahn, "Digital Equalization of Chromatic Dispersion and Polarization Mode Dispersion," J. Lightwave Technol., **25**, 2033-2043 (2007).

[12] G. Goldfarb and G. Li, "Efficient backward-propagation using wavelet based filtering for fiber backward-propagation," Opt. Exp., **17**, 8815-8821 (2008).

[13] Mats Sköld, Mathias Westlund, Henrik Sunnerud, and Peter A. Andrekson, "All-Optical Waveform Sampling in High-Speed Optical Communication Systems Using Advanced Modulation Formats," J. Lightwave Technol. **27**, 3662-3671 (2009)

[14] S. Gupta, D. R. Solli , A. Motafakker-Fard, and B. Jalali, "Capturing Rogue Events with the Time-Stretch Recording Scope," in CLEO/Europe-EQEC Conference, (2009), paper CF6.1.

[15] Y. Han, O. Boyraz, and B. Jalali, "Ultrawide-band photonic time-stretch A/D converter employing phase diversity," IEEE trans. Microwave theory and techniques, **53**, pp. 1404-1408.

[16] Y. Han and B. Jalali, "Photonic Time-Stretched Analog-to-Digital Converter: Fundamental Concepts and Practical Considerations," J. Lightwave Technol., **21**, No.12, 3085-3103, (2003).

[17] S. Gupta, S.; Jalali, B.; Stigwall, J.; Gait, S., "Demonstration of Distortion Suppression in Photonic Time-Stretch ADC using Back Propagation Method," in 2007 IEEE International Topical Meeting on Microwave Photonics , pp.141-144, 3-5 Oct. 2007.

[18] "GigOptix`s 100G Mach Zehnder Modulator Enables 110GHz Time Stretched Analog to Digital Conversions at UCLA"  www.reuters.com, Aug. 11, 2009.

[19] H. Bülow, F. Buchali, and A. Klekamp, "Electronic Dispersion Compensation," J. Lightwave Technol. **26**, 158-167 (2008).